\documentclass[a4paper,11pt]{article}
\usepackage{jinstpub} 
\usepackage{lineno}
\usepackage{siunitx}
\usepackage[tight]{subfigure}
\usepackage{wrapfig}

\graphicspath{{Figures/}}

\title{\boldmath Measurement of scintillation from proportional electron multiplication in liquid xenon using a needle}

\author[a,1]{P.~Knights,\note{Corresponding author.}}
\emailAdd{p.r.knights@bham.ac.uk}
\author[b,c]{H.~Sekiya,}
\author[d]{I.~Katsioulas,}
\author[a,e]{K.~Nikolopoulos,}
\author[f]{K.~Kanzawa,}
\author[g]{and I.~Giomataris}

\affiliation[a]{School of Physics and Astronomy, University of Birmingham,\\
Edgbaston, B15 2TT, Birmingham, United Kingdom}
\affiliation[b]{Kamioka Observatory, Institute for Cosmic Ray Research, The University of Tokyo,\\
456 Higashi-Mizumi, Kamioka, Hida, Gifu, 506-1205, Japan}
\affiliation[c]{Kavli Institute for the Physics and Mathematics of the Universe,The University of Tokyo\\
5-1-5 Kashiwanoha, Kashiwa, Chiba, 277-8583, Japan}
\affiliation[d]{European Spallation Source, PO Box 176, 22100, Lund, Sweden}
\affiliation[e]{Institute of Experimental Physics, University of Hamburg,\\ Luruper Chaussee 149, 22761, Hamburg, Germany}
\affiliation[f]{Institute for Space-Earth Environmental Research, Nagoya University,\\ 
Furo-cho, Chikusa-ku, Nagoya 464-8601, Japan }
\affiliation[g]{IRFU, CEA, Universit\'e Paris-Saclay, F-91191 Gif-sur-Yvette, France}

\abstract{Charge amplification in liquids could provide single-phase xenon time projection chambers with background discrimination and fiducialisation capabilities similar to those found in dual-phase detectors. Although efforts to achieve the high electric field required for charge amplification and proportional scintillation in liquid xenon have been previously reported, their application to large-scale detectors remains elusive. This work presents a new approach to this challenge, where -- instead of the  thin-wire approach of previous studies -- a needle-like high-voltage electrode is employed to demonstrate proportional charge amplification and secondary scintillation production in liquid xenon. This is an important milestone towards the development of an electrode structure that could be utilised in a large-scale, single-phase time projection chamber with dual read-out.
}


\begin{document}
\maketitle
\flushbottom

\section{Introduction}
\label{sec:intro}
A wealth of cosmological and astrophysical evidence indicates the presence of an as-of-yet unknown mass component of the universe, named Dark Matter (DM). While increasingly more sensitive direct searches for DM have been performed over a number of decades, no conclusive observation has been made~\cite{Billard:2021uyg}. Most experimental efforts in this regard have focused on the detection of nuclear recoils induced by scatterring of DM candidates with a mass in the region between $10$ and $1000\;\si{\giga\eV}$, corresponding to the expected mass of a canonical thermal Weakly Interacting Massive Particle~\cite{10.1093/ptep/ptac097}.
Current world-leading sensitivities in this mass region have been achieved by experiments employing dual-phase time projection chambers (TPCs) filled with noble elements, primarily xenon and argon. 

Dual-phase TPCs operate by detecting scintillation light produced by the initial particle interaction in the liquid, referred to as the S1 signal, and the secondary electroluminescence 
produced by the multiplication of ionisation electrons extracted into the gas phase, referred to as the S2 signal. 
The combination of these two signals provides invaluable fiducialisation and event identification capabilities. The position of S2, inferred using a segmented light read-out plane, along with the time difference between S1 and S2, provide the initial interaction location. Additionally, the relative strengths of S1 and S2 are different for electronic and nuclear recoils, allowing for background discrimination.

While this technology has proved successful in improving sensitivity by many orders of magnitude in a few decades, several technical challenges mean that direct scaling of the technology to ever larger detectors is demanding~\cite{Aprile:2014ila}.
One method to overcome these challenges is the use of a single-phase TPC~\cite{Kuger:2022}, where the absence of a gas-liquid interface reduces reflections and, thus, increases the S1 light yield. However, producing an S2 signal through the amplification of the ionisation electrons directly in the liquid is a major challenge to the full realisation of this concept.
Previous efforts have demonstrated the feasibility of generating an S2 in liquid xenon, using thin, $\mathcal{O}(10\;\si{\micro\meter})$ wires~\cite{LANSIART197647, Masuda:1978tjp,Aprile:1992de,BENETTI1993203,Aprile:2014ila, Qi:2023bof} or gas electron multipliers (GEMs)~\cite{Arazi:2015prw}, to produce the electric field required for amplification. The technical challenge of incorporating such amplification structures into a single-phase TPC remains. 

Another advantage of the single-phase implementation over the dual-phase TPC is the ability to use a spherical design, increasing further the intrinsic light yield by tiling the spherical surface with photo-detection modules. This approach was employed by the XMASS collaboration~\cite{Abe:PhysRevD.108.083022,Abe:NIM2013}, utilising an almost spherical detector lined with photomultiplier tubes (PMTs). However, XMASS only observed the S1 signal with no charge amplification to generate an S2. To transform an XMASS-like detector into a true single-phase TPC, while preserving the spherical symmetry, a different charge amplification structure to those already demonstrated in liquid xenon would be required. 

The spherical proportional counter (SPC)~\cite{Giomataris:2008ap, Arnaud:2018bpc, Katsioulas:2018pyh, Giomataris:2020rna}, similarly to XMASS, is a spherical detector; however, it operates in the gas phase and reads out the ionisation signal from a central electrode. The SPC is used by the NEWS-G collaboration for DM searches~\cite{NEWS-G:2017pxg, NEWS-G:2021vfh, NEWS-G:2023qwh}. The detector comprises one or more spherical anodes, with $\mathcal{O}(\si{\milli\meter})$ diameter, placed at the centre of a grounded spherical cathode, with the detector volume filled with gas.
Advances in the SPC central electrode instrumentation have enabled increased gas gain, that is electron multiplication factors, at lower voltages, enhanced detector operation stability, and improved response uniformity.
~\cite{Katsioulas:2018pyh, Giomataris:2020rna}. Such an electrode, particularly the multi-anode ACHINOS structure~\cite{Giganon:2017isb, Giomataris:2020rna, Herd:2023hmu}, is a promising option for charge amplification in a spherical single-phase TPC. However, to achieve the required electric fields for amplification in the liquid phase without drastically increasing the voltage and jeopardising detector stability, sharper amplification structures are required. 

In this article the first endeavour in this direction is presented, using a thin needle-like structure to generate charge amplification, and an S2 signal, in a single-phase liquid xenon TPC test bench. In Section~\ref{sec:setup} the employed amplification structure is presented, along with the experimental set-up, while in Section~\ref{sec:results} the measurements performed are presented and discussed.

\section{Experimental Setup}
\label{sec:setup}
\begin{figure}[b]
\centering
    \subfigure[\label{fig:tpcSetup}]{\includegraphics[width=0.40\linewidth]{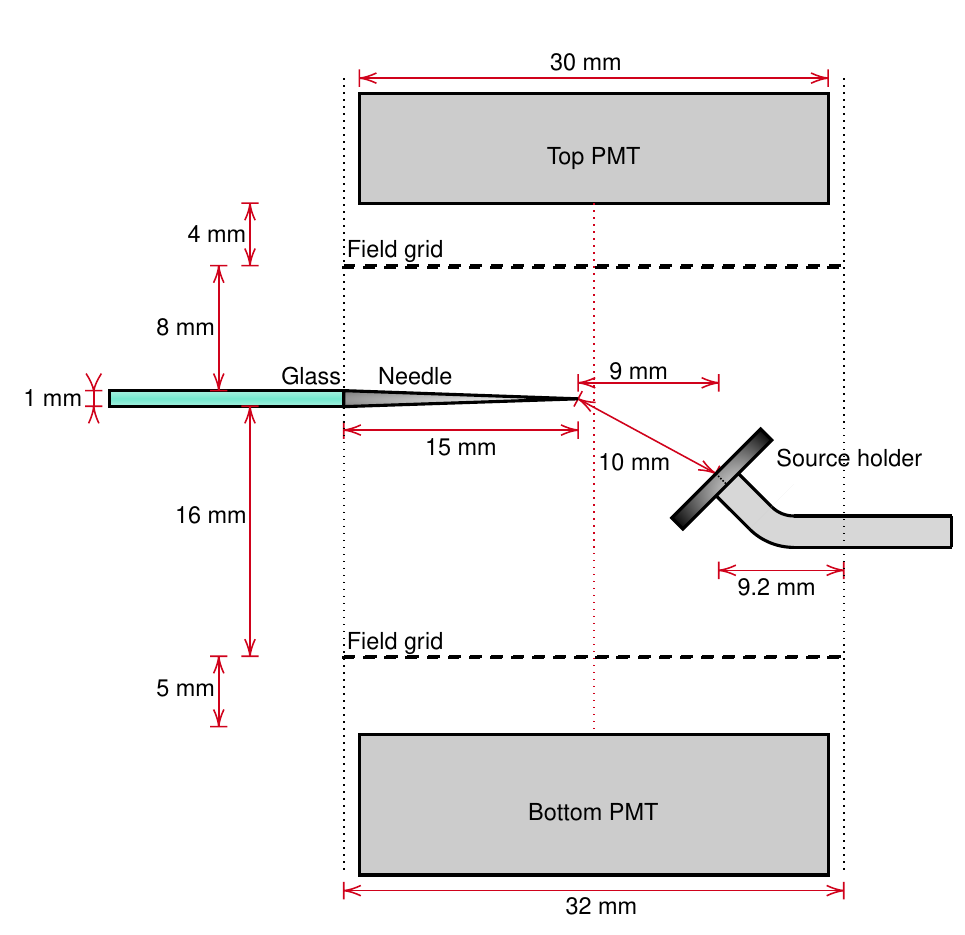}}%
    \subfigure[\label{fig:TheSensor}]{\includegraphics[width=0.48\linewidth]{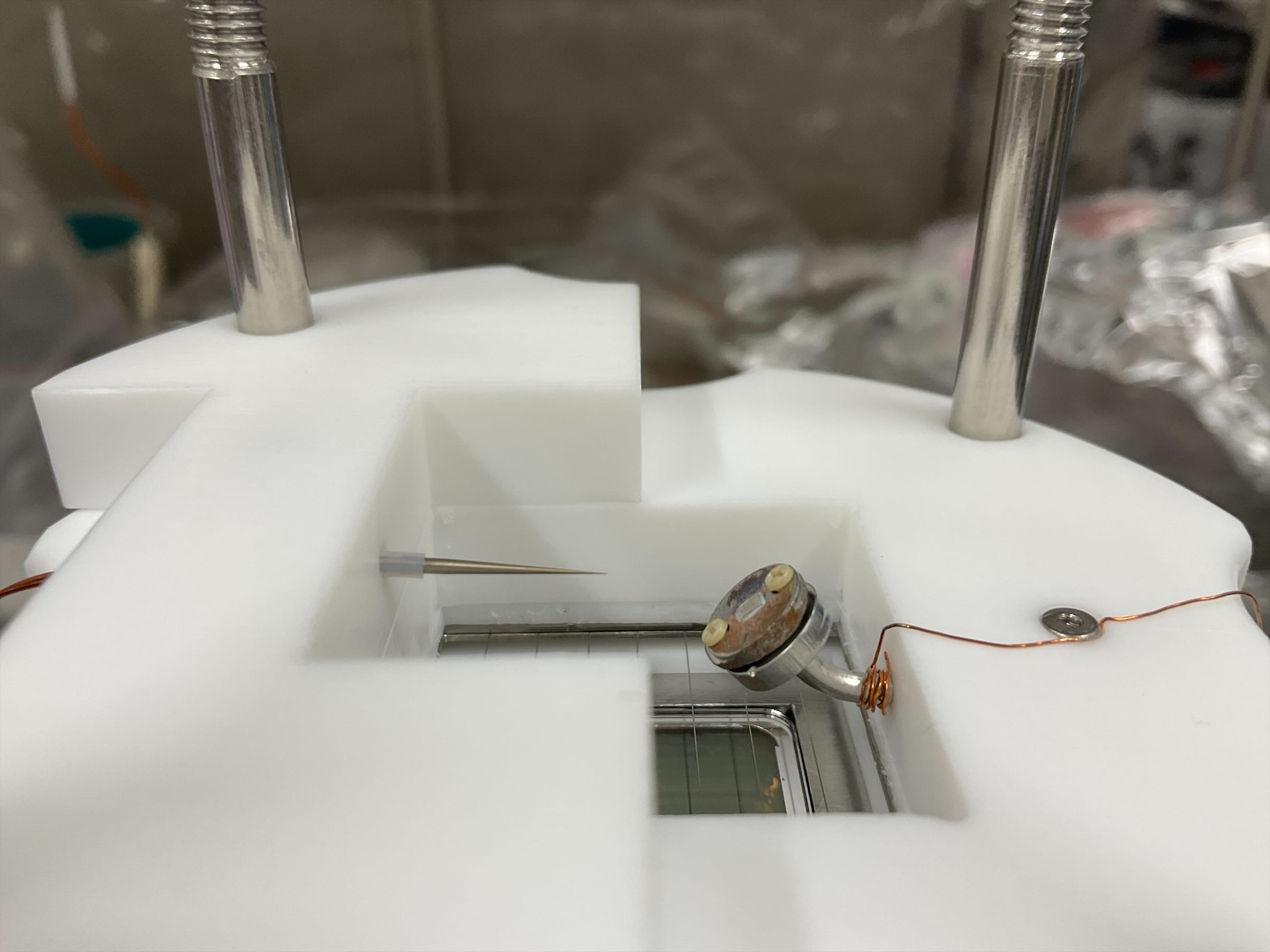}}%
    \caption[]{\subref{fig:tpcSetup} Experimental set-up showing the TPC region, contained within the two field grids and observed by the Top and Bottom PMTs. All components are held in place with polytetrafluoroethylene support structures, except the PMTs which are held by stainless steel brackets, which are not shown for clarity. \subref{fig:TheSensor} Photograph of the inside of the TPC, showing the needle and source holder. \label{fig:sensorAndSource}} 
\end{figure}
\begin{figure}[htbp] 
  \centering
\subfigure[\label{fig:daq}]{\includegraphics[width=0.56\linewidth]{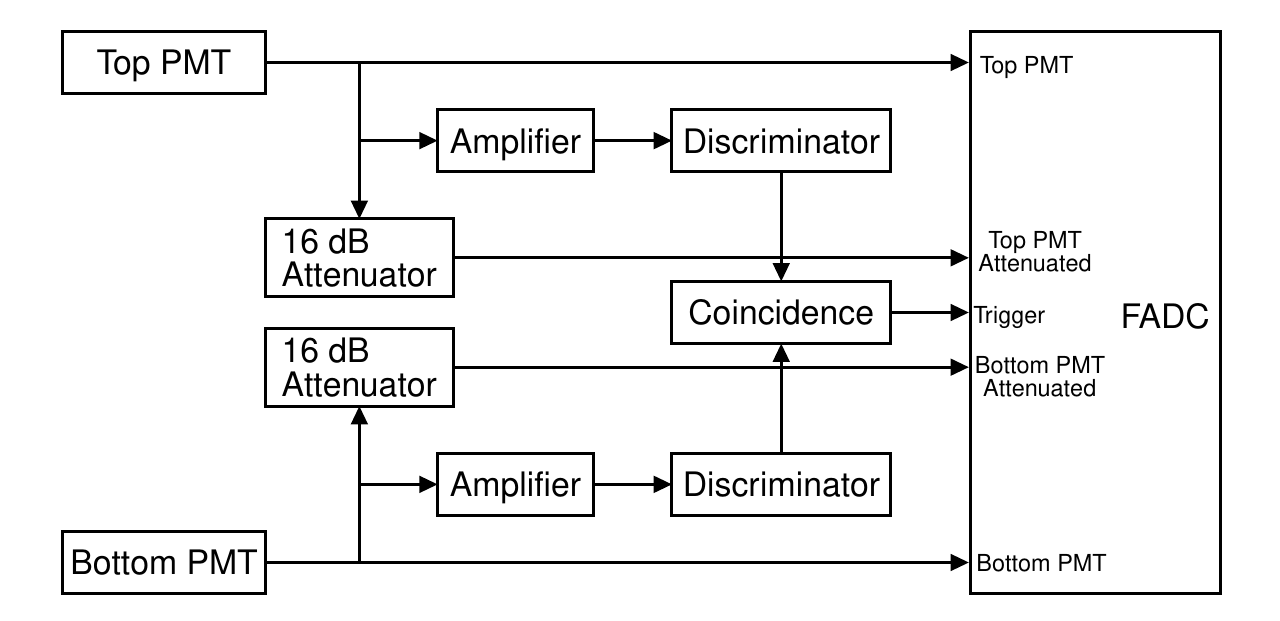}}
\subfigure[\label{fig:fullExperiment}]{\includegraphics[width=0.32\linewidth]{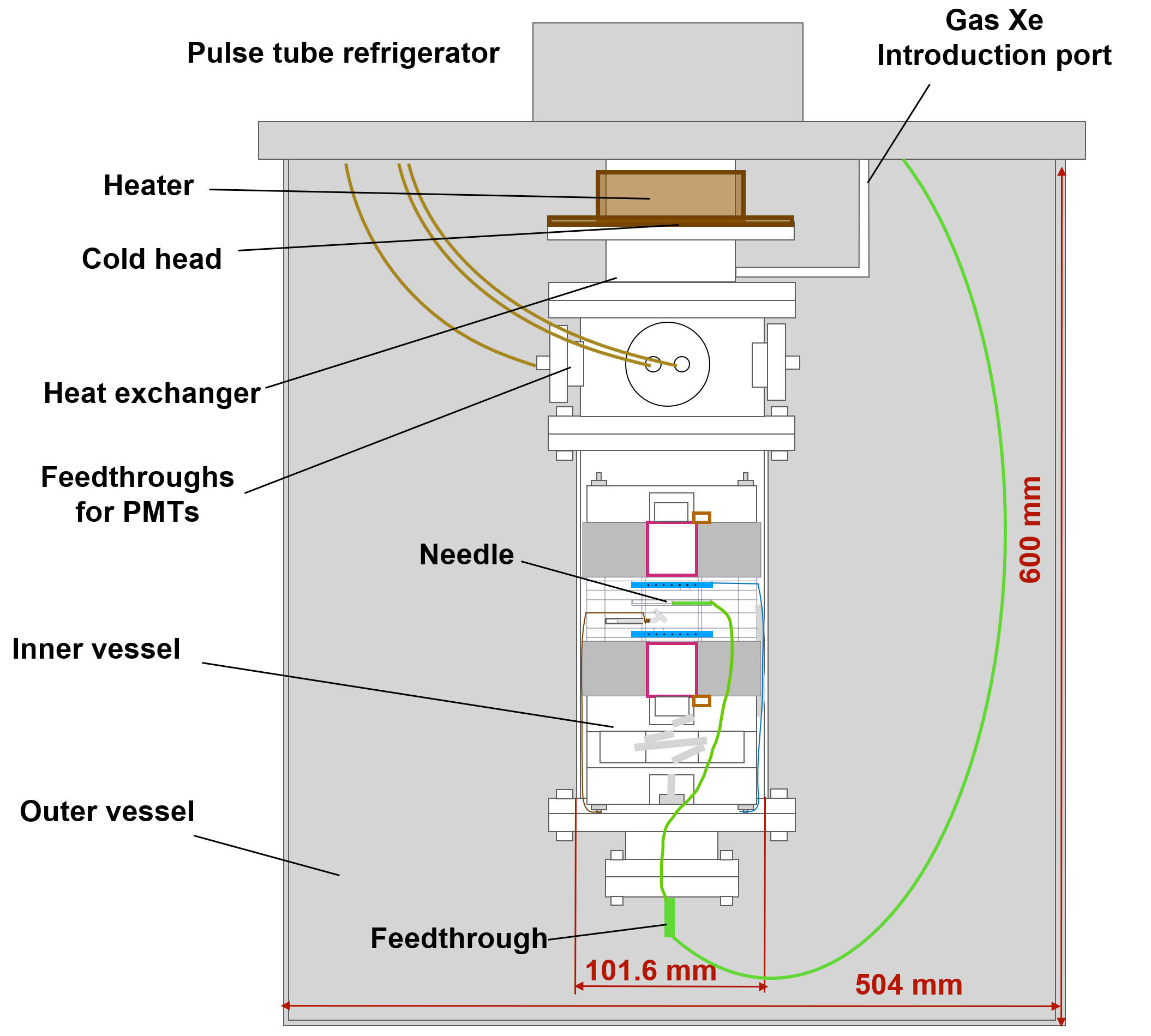}}
  \caption{
  \subref{fig:daq} Data acquisition chain. 
  \subref{fig:fullExperiment} The experimental set-up
  }
\end{figure}

The configuration of the single-phase xenon TPC test bench is shown in Figure~\ref{fig:tpcSetup}. The test bench comprises two Hamamatsu R8520-406 PMTs, mounted above and below two field cage meshes, which define the active region. The field cages consist of a grounded woven-tungsten mesh made of 100~$\mu$m diameter wire. In the active region, a source holder, shown in Figure~\ref{fig:TheSensor}, supports a $^{241}$Am source with a $2.0\times1.5\;\si{\milli\meter}^2$ active area. This provides a source of $\alpha$-particles, typically with an energy of approximately $5.5\;\si{\mega\eV}$, and a series of X-rays with energy below approximately $60\;\si{\kilo\eV}$. The source is mounted opposite to the charge amplification electrode and the tip of the electrode is located $10\;\si{\milli\meter}$ from the source. All components are held in place by a series of PTFE components, visible in Figure~\ref{fig:TheSensor}, inside a $10\;\si{\centi\meter}$ in diameter cylindrical chamber, referred to as the inner vessel. A $250\;\si{\micro\meter}$ ceramic-insulated copper wire connects the electrode to an SHV feed-through at the bottom of the chamber. In the final design, the feed-through was located at the bottom of chamber, shown in Figure~\ref{fig:fullExperiment}, ensuring it was submerged in the liquid xenon to suppress electric discharges.

\begin{wrapfigure}{L}{0.45\linewidth}
\vspace{-0.5cm}
\centering
\includegraphics[width=.99\linewidth]{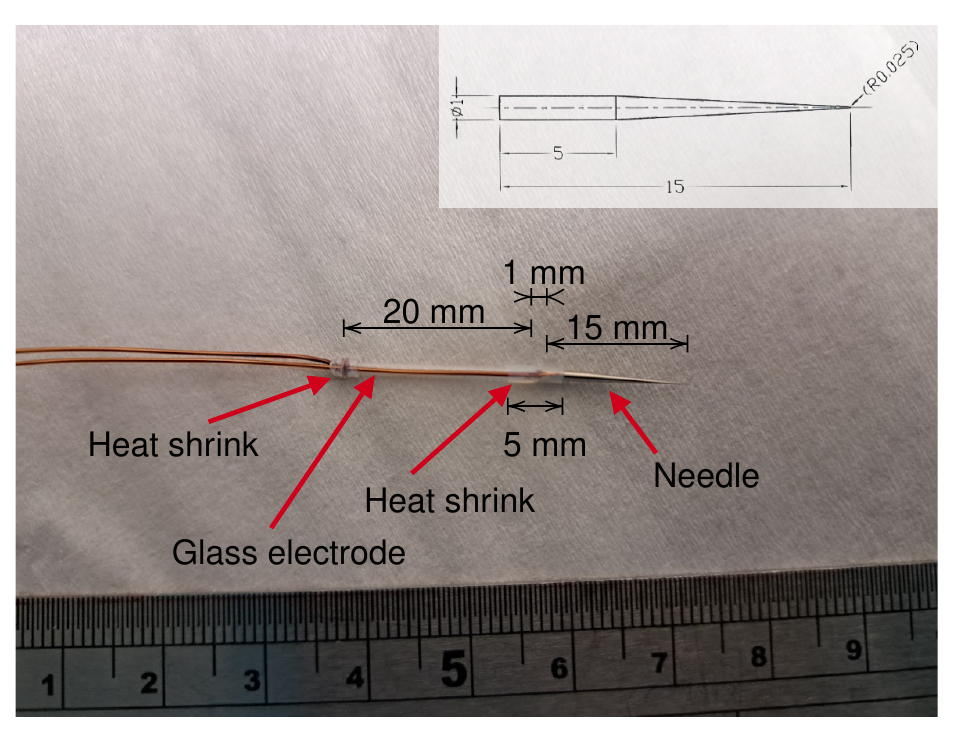}
\caption{The needle electrode and its support structure, featuring a grounded, resisitve glass electrode. The inset shows the needle design. \label{fig:sensor}}
\end{wrapfigure}
The needle electrode used as the basis for the charge amplification structure is shown in Figure~\ref{fig:sensor}.
Needles with $50\;\si{\micro\meter}$ in diameter tips were employed, to reduce the voltage required for amplification compared to earlier test. To hold the needle in place, a grounded, resistive electrode made of a soda-lime glass tube, $1\;\si{\milli\meter}$ in external diameter, was used, similar to that discussed in Ref.~\cite{Katsioulas:2018pyh}. The use of a resistive material to support the needle reduced the potential for discharge or charging up, compared to a conductive or insulator, respectively. The wire from the electrode passed through the inside of the tube, while the tube itself was connected to a second wire and grounded. This is shown in Figure~\ref{fig:sensor}.

The data acquisition chain is shown in Figure~\ref{fig:daq}. A four-channel flash analog-to-digital converter (FADC) (CAEN, DT5751) with  a dynamic range of $1\;\si{\volt}$, and a sampling rate of $1\;\si{\giga\hertz}$ is used.
To capture changes in the gain of the $5.5\;\si{\mega\eV}$ $\alpha$ signal within the dynamic range of the FADC, the signal outputs from the top and bottom PMTs are divided into two parts using dividers, one part was input directly to the FADC, and the other part was attenuated by 16 dB and input to the FADC. The PM-amp is used to increase the wave height of the signals from the upper and lower PMTs by a factor of 10, and the signals that exceeded the threshold value ($55\;\si{\milli\volt}$) by the discriminator are used as the trigger signal for the FADC.

The system to support and cool the inner vessel is shown in Figure~\ref{fig:fullExperiment}.
During operation with liquid xenon, the inner vessel was cooled by a pulse tube refrigerator (ULVAC Cryo PC150U+UW701) mounted to the top of the vessel. The temperature was regulated to a constant $-113\si{\degreeCelsius}$ by a current-heated wire also mounted to the top of the vessel. The temperature in the system was continuously monitored at several points throughout the inner vessel. 
The inner vessel was contained within a cylindrical outer vacuum vessel, $50.4\;\si{\centi\meter}$ in diameter and $60\;\si{\centi\meter}$ in height, to thermally isolate it from the room. During operation, the outer vacuum vessel was continuously evacuated using a turbo-molecular pump
and maintained at a pressure of $7\times10^{-3}\;\si{\milli\bar}$.

\begin{figure}[b]
    \centering
    \subfigure[\label{fig:gasSystem}]{\includegraphics[width=0.57\linewidth]{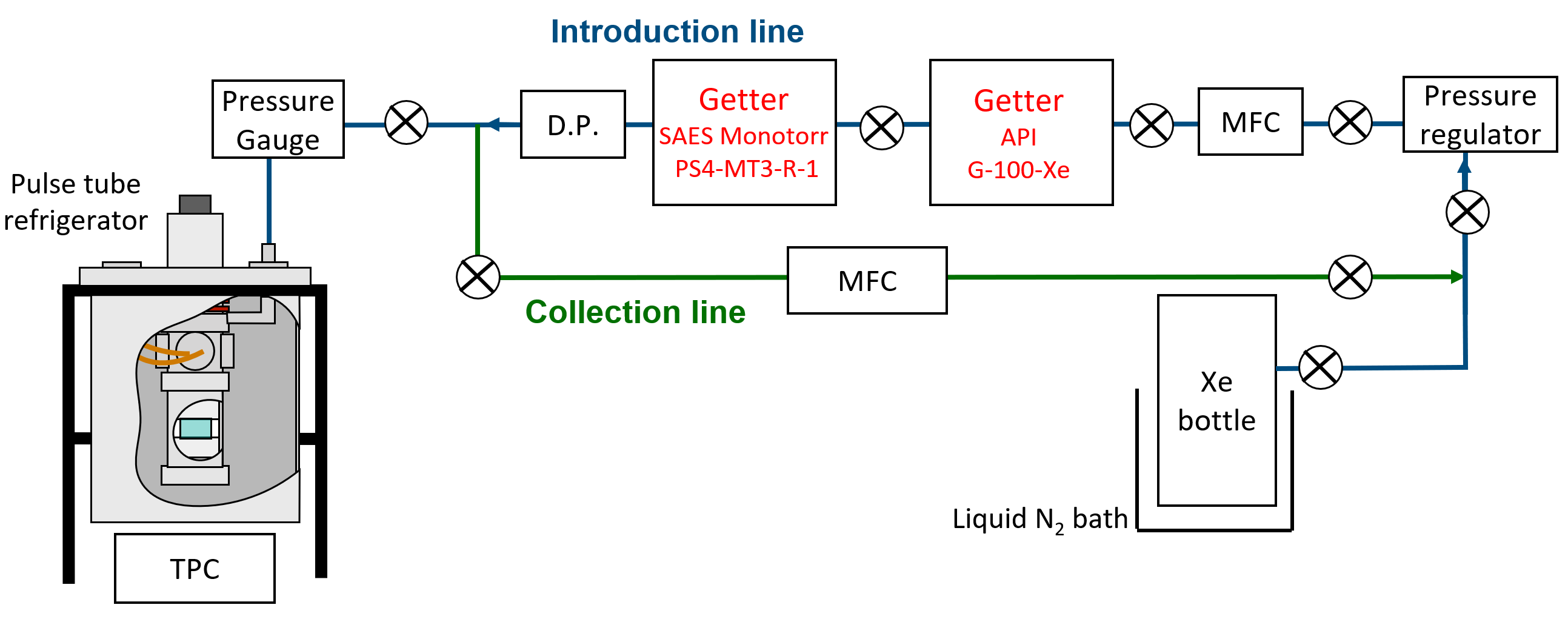}}
    \subfigure[\label{fig:temperature}]{\includegraphics[width=0.42\linewidth]{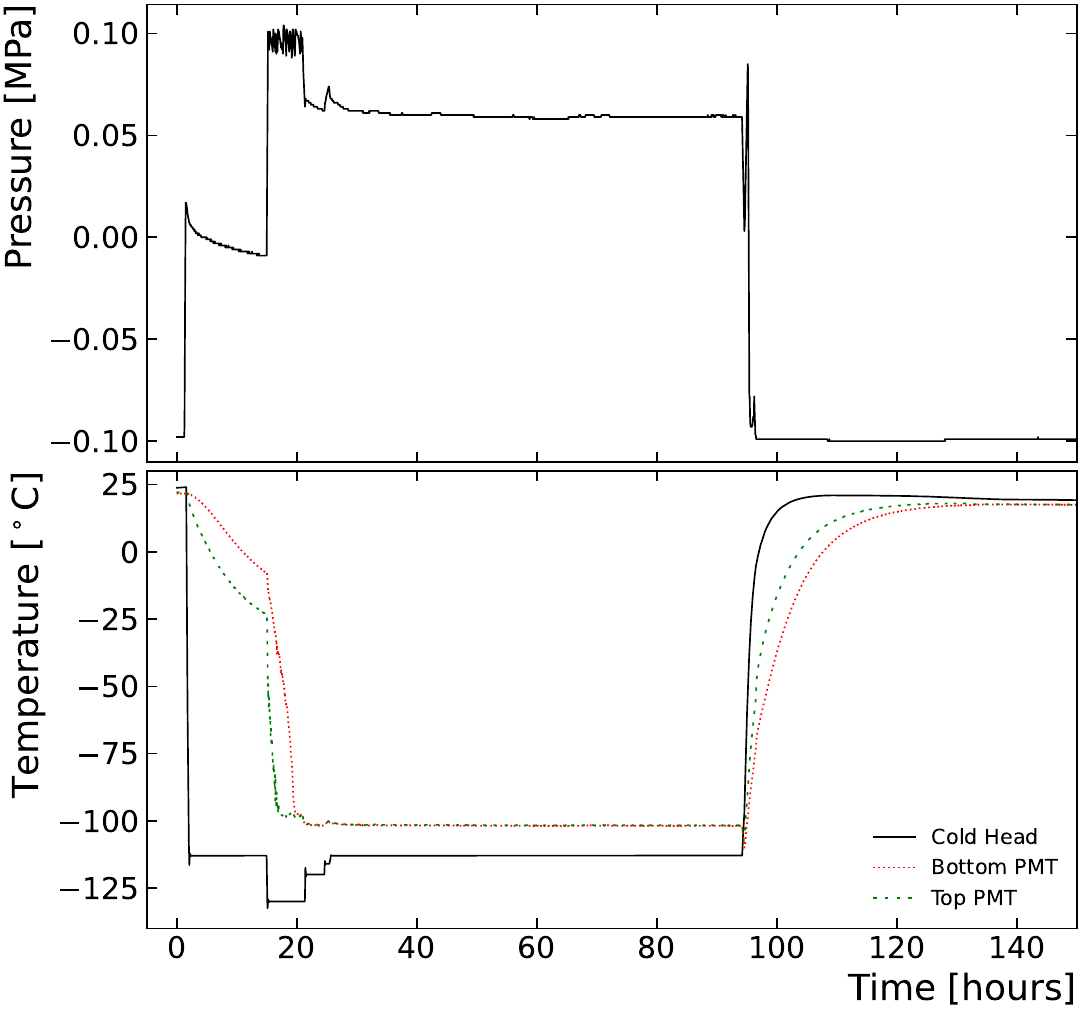}}    
\caption{\subref{fig:gasSystem} The Xe handling system.
\subref{fig:temperature} Temperature and pressure during the cooling, measurement, and xenon collection
    \label{fig:xenonSystem}}
\end{figure}

The xenon handling system is shown schematically Figure~\ref{fig:gasSystem}. Prior to filling the inner vessel with xenon, it was evacuated over several days to a vacuum of $\num{4.5e-7}\;\si{\milli\bar}$. The xenon was introduced into the inner vessel via a SAES Monotorr (PS4-MT3-R-1) and API G-100-Xe purifier to remove residual oxygen and water contamination, and finally, through a $3\;\si{\nano\meter}$ particulate filter.
The dew point of the xenon was monitored via a Michel Instrument Pura dewpoint meter and was kept $<-100^\circ$C. 
The precooling and cooling of the vessel took place over the course of $24\;\si{hours}$.  The history of the temperature and the pressure during this campaign is shown in Figure~\ref{fig:temperature}.

\section{Tests in Liquid Xenon}
\label{sec:results}
\begin{wrapfigure}{L}{0.4\linewidth}
\vspace{-0.5cm}
\centering
\includegraphics[width=.99\linewidth]{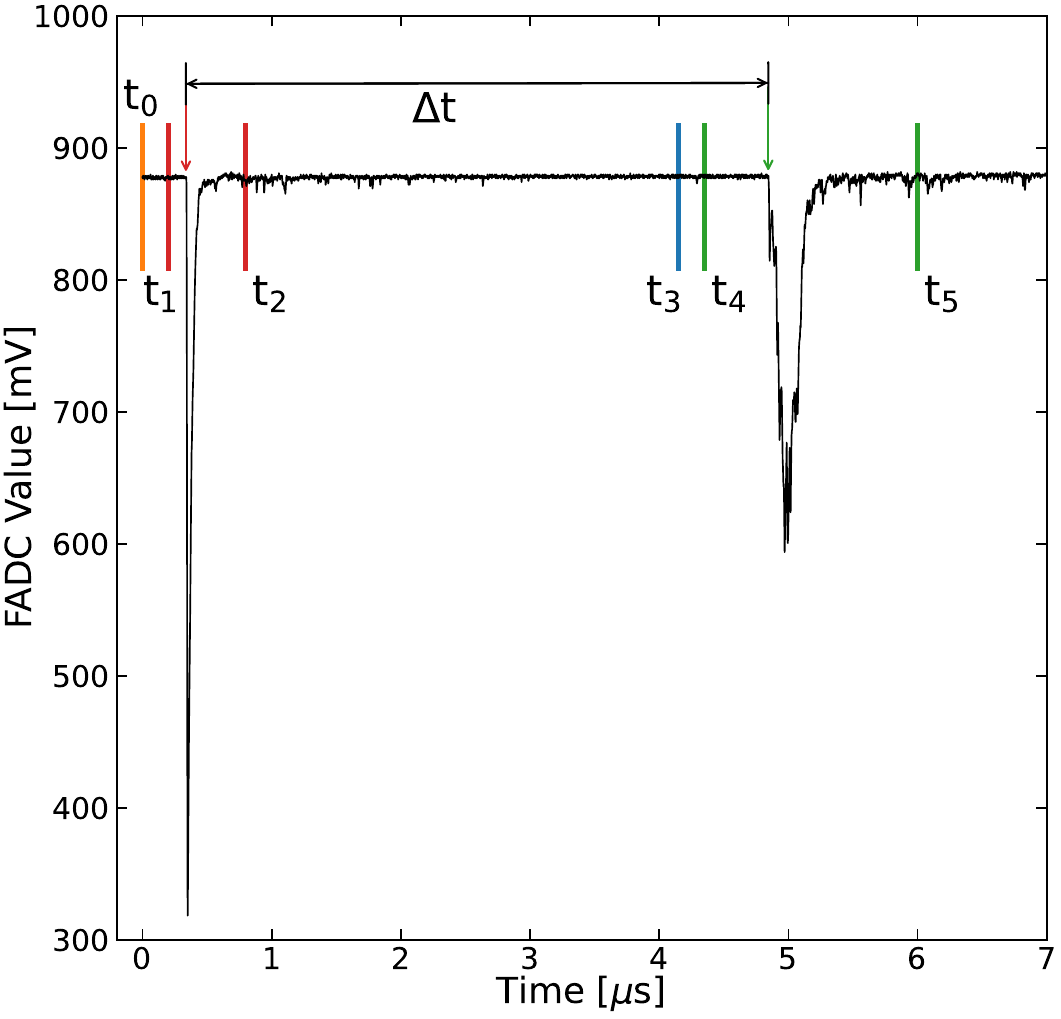}
\vspace{-0.75cm}
\caption{Annotated recorded trace showing the time difference $\Delta t$ between S1 and S2. The S1 integration window is between $t_{1}$ and $t_{2}$, with the baseline computed between $t_{0}$ and $t_{1}$. For S2, the baseline is computed between $t_{3}$ and $t_{4}$, and the signal integrated between $t_{4}$ and $t_{5}$. \label{fig:annotatedEvent}}
\vspace{-0.6cm}
\end{wrapfigure}
The top and bottom PMTs were biased to $792\;\si{\volt}$ and $892\;\si{\volt}$, respectively adjusting their gains to match at $7\times10^6$. An example recorded trace is shown in Figure~\ref{fig:annotatedEvent}, along with the parameters used for the analysis. The baseline, defined as the average FADC value $200\;\si{\nano\second}$ before the S1 trigger was computed. The area of the S1 signal was computed by integrating between $200\;\si{\nano\second}$ and $800\;\si{\nano\second}$ of the trace and subtracting the baseline integral over this range. 
The onset time for the S1 signal was computed as the time at which the signal drops to less than $1\%$ below the baseline value immediately before the lowest point of the S1 signal. A similar process was used in the expected S2 region, computing the baseline between $4.05\;\si{\micro\second}$ and $4.25\;\si{\micro\second}$, and integrating the area to $6.0\;\si{\micro\second}$. The time between the S1 and S2 signals $\Delta t$ is defined as the difference between the two onset times.

\begin{figure}[b]
\centering
\includegraphics[width=.9\textwidth]{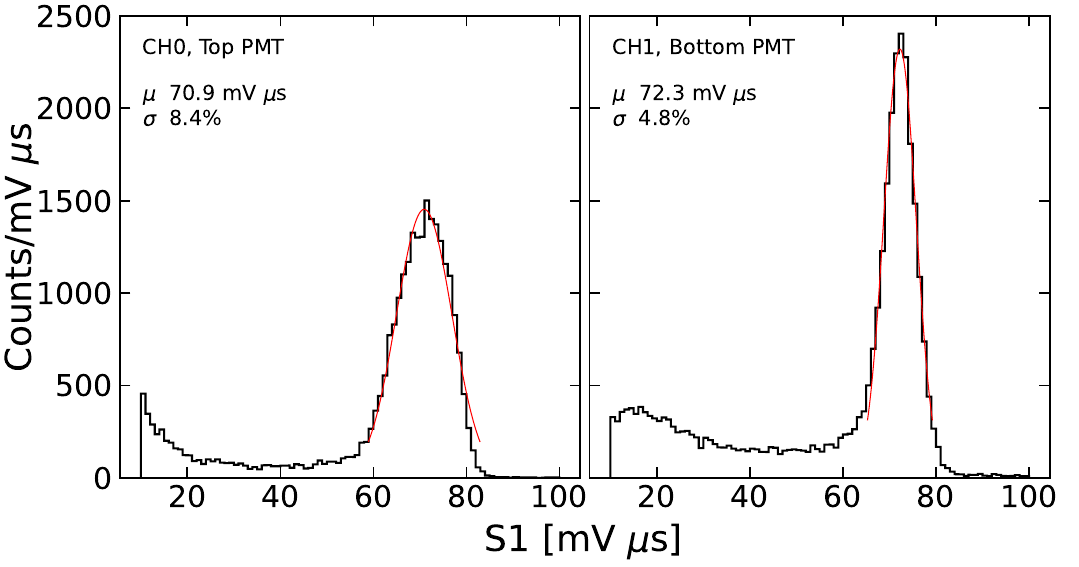}
\caption{Example S1 distribution for the Top and Bottom PMTs. The peak corresponds to the $^{241}$Am $\alpha$-particle, and is fit with a Gaussian function to extract the mean value.\label{fig:s1distribution}}
\end{figure}
\begin{wrapfigure}{L}{0.4\linewidth}
\vspace{-0.5cm}
\centering
\includegraphics[width=0.99\linewidth]{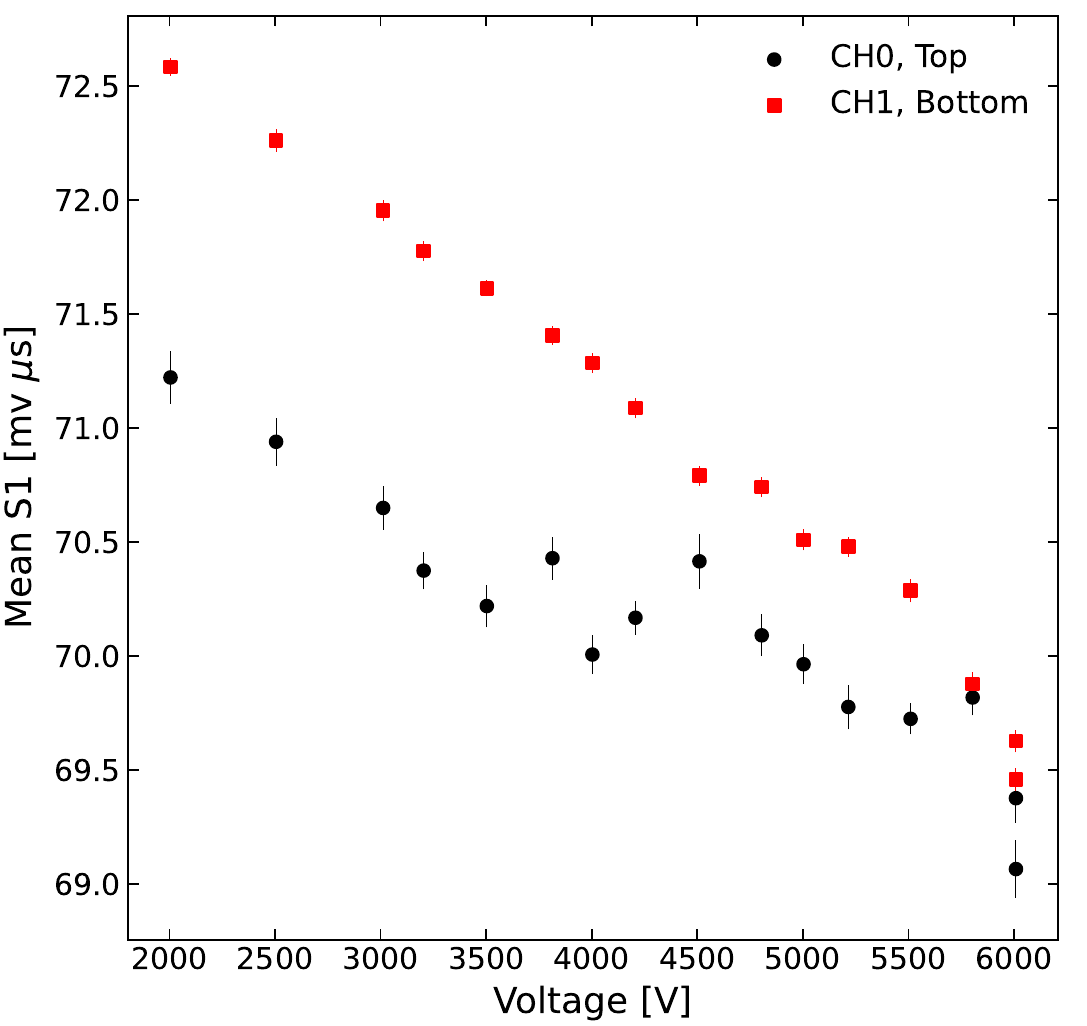}
\vspace{-0.75cm}
\caption{Mean S1 as a function of voltage applied to the needle. \label{fig:s1WithVoltage}}
\vspace{-0.5cm}
\end{wrapfigure}

An example of S1 signal recorded 
from each channel when the needle had no voltage applied is shown in Figure~\ref{fig:s1distribution}. The 
peak at approximately $70\;\si{\milli\volt\micro\second}$ corresponds to the full energy of the $\alpha$-particle from the $^{241}$Am source. The low energy tail is attributed to 
the distribution of $\alpha$-particle energies after emerging from some depth within the source, as well as to Compton scatterings in the xenon of photons  emitted from the source. The peak was fit with a Gaussian to extract its position, which was used to monitor the change in S1 with voltage. The voltage applied to the needle was varied between $0\;\si{\volt}$ and $6000\;\si{\volt}$, and the S1 signal as a function of voltage is shown in Figure~\ref{fig:s1WithVoltage}. As expected, there is a slight decrease in S1 with voltage, attributed to a reduction in the recombination of ionisation electrons as the collection electric field increases.

\begin{figure}[b] 
  \centering
\subfigure[\label{subfig:exampleEvent2500V}]{\includegraphics[width=0.32\linewidth]{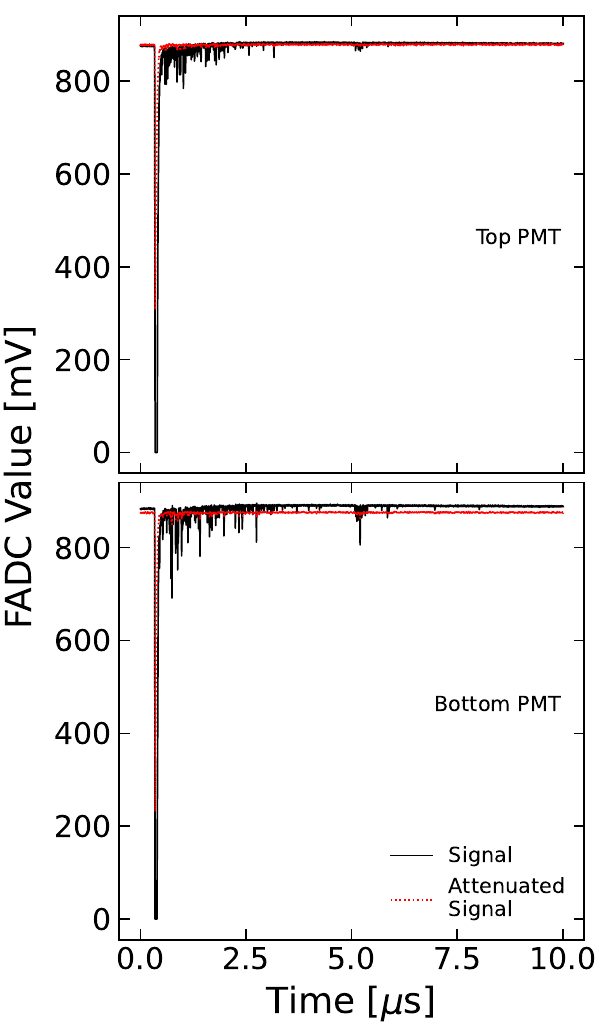}}
\subfigure[\label{subfig:exampleEvent4500V}]{\includegraphics[width=0.32\linewidth]{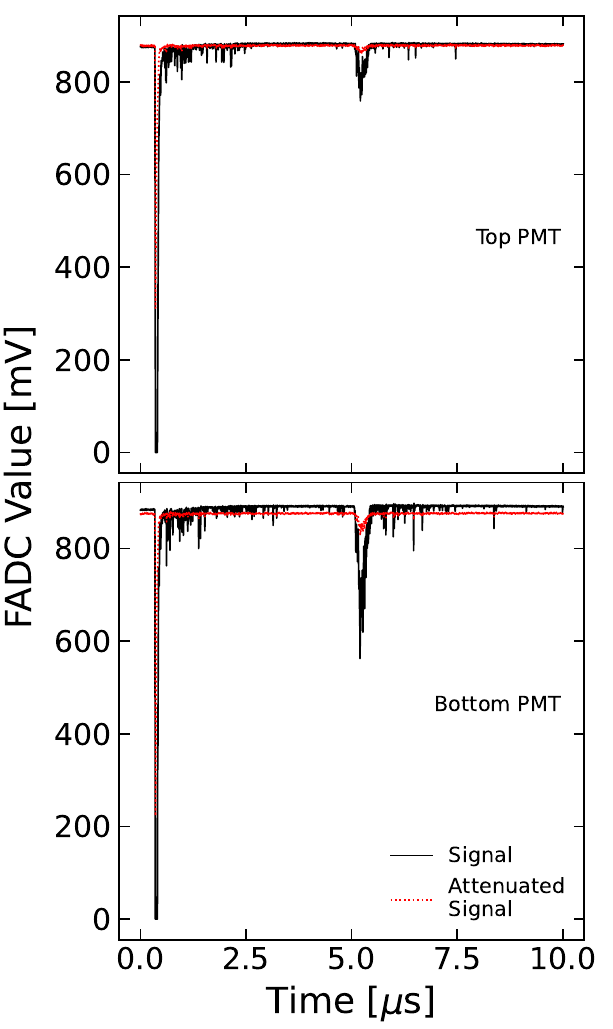}}
\subfigure[\label{subfig:exampleEvent6000V}]{\includegraphics[width=0.32\linewidth]{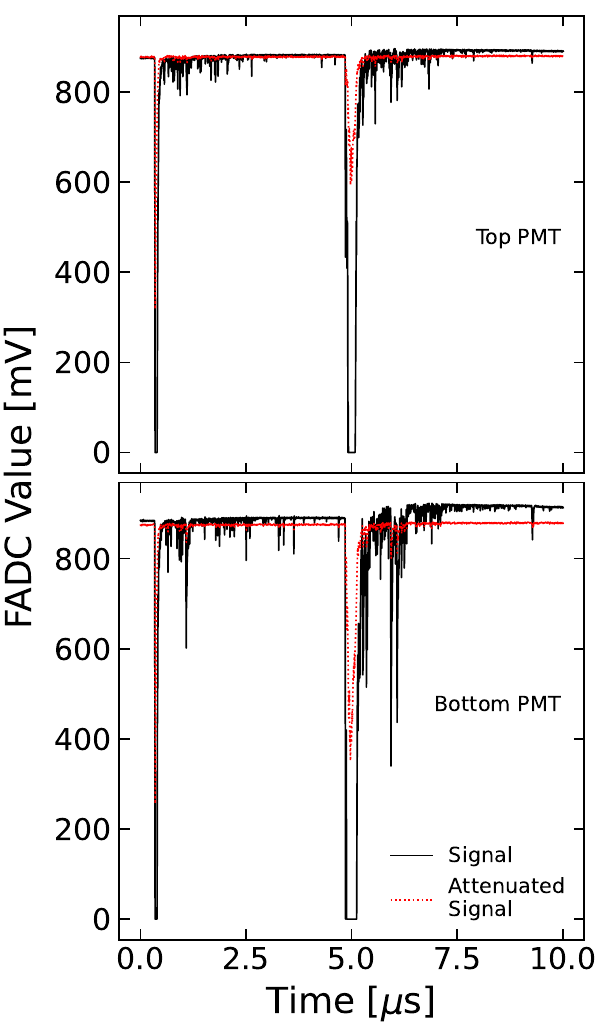}}
  \caption{Example recorded signals for \subref{subfig:exampleEvent2500V} $2505\;\si{\volt}$, \subref{subfig:exampleEvent4500V} $4510\;\si{\volt}$, and \subref{subfig:exampleEvent6000V} $6009\;\si{\volt}$ applied to the needle.\label{fig:exampleTraces}}
\end{figure}

For voltages above $2000\;\si{\volt}$, an S2 signal was observed in the recorded PMT signals. Figure~\ref{fig:exampleTraces} shows examples of the recorded signals at $2505\;\si{\volt}$, $4510\;\si{\volt}$ and $6009\;\si{\volt}$ applied to the needle, where S2 clearly increases with applied voltage. 

\begin{figure}[h] 
  \centering
\subfigure[\label{subfig:s22500}]{\includegraphics[width=0.32\linewidth]{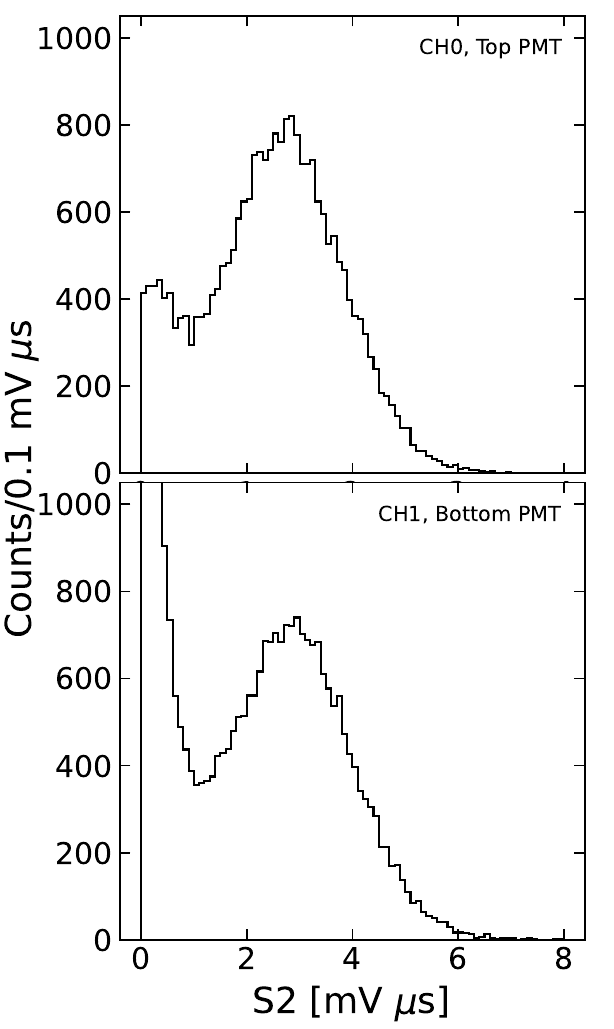}}
\subfigure[\label{subfig:s24500}]{\includegraphics[width=0.32\linewidth]{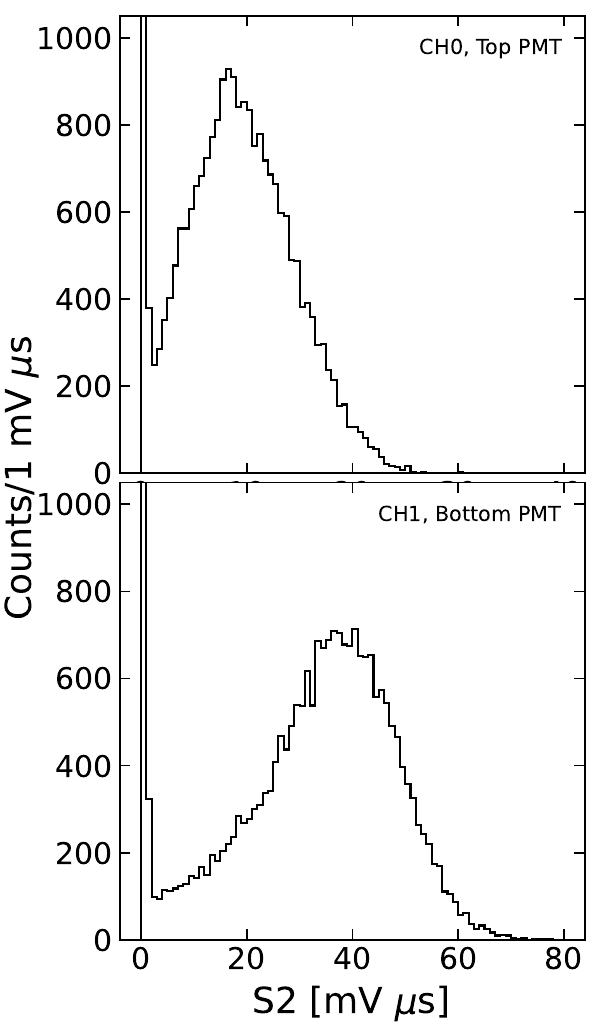}}
  \subfigure[\label{subfig:s26000}]{\includegraphics[width=0.32\linewidth]{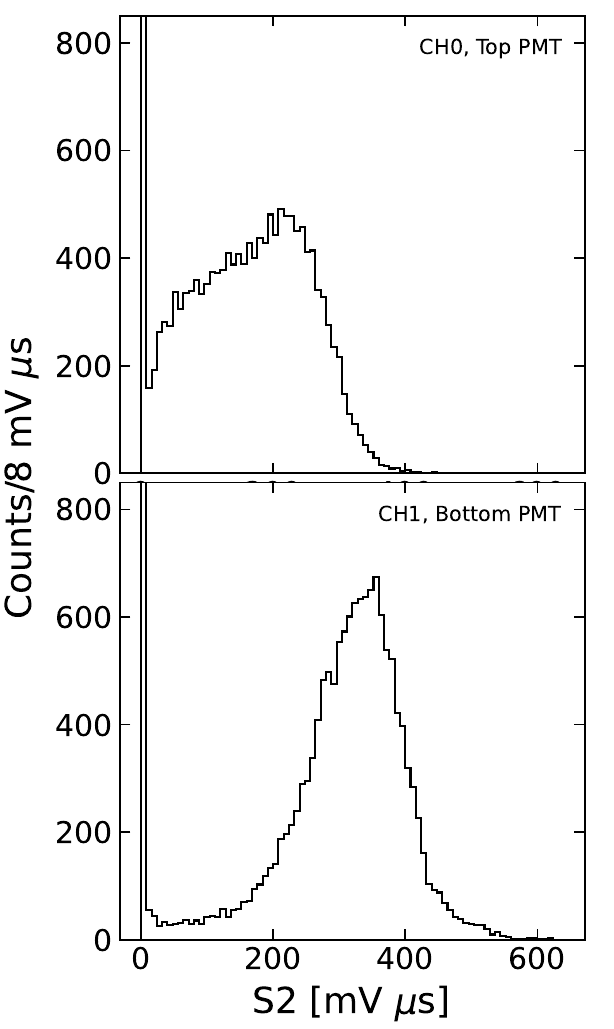}}
  \caption{\label{fig:s2Areas}Integrated S2 signal when \subref{subfig:s22500} $2505\;\si{\volt}$, \subref{subfig:s24500} $4510\;\si{\volt}$, and \subref{subfig:s26000} $6009\;\si{\volt}$ was applied to the needle.}
\end{figure}

\begin{figure}[htbp]
\centering
\includegraphics[width=.49\textwidth]{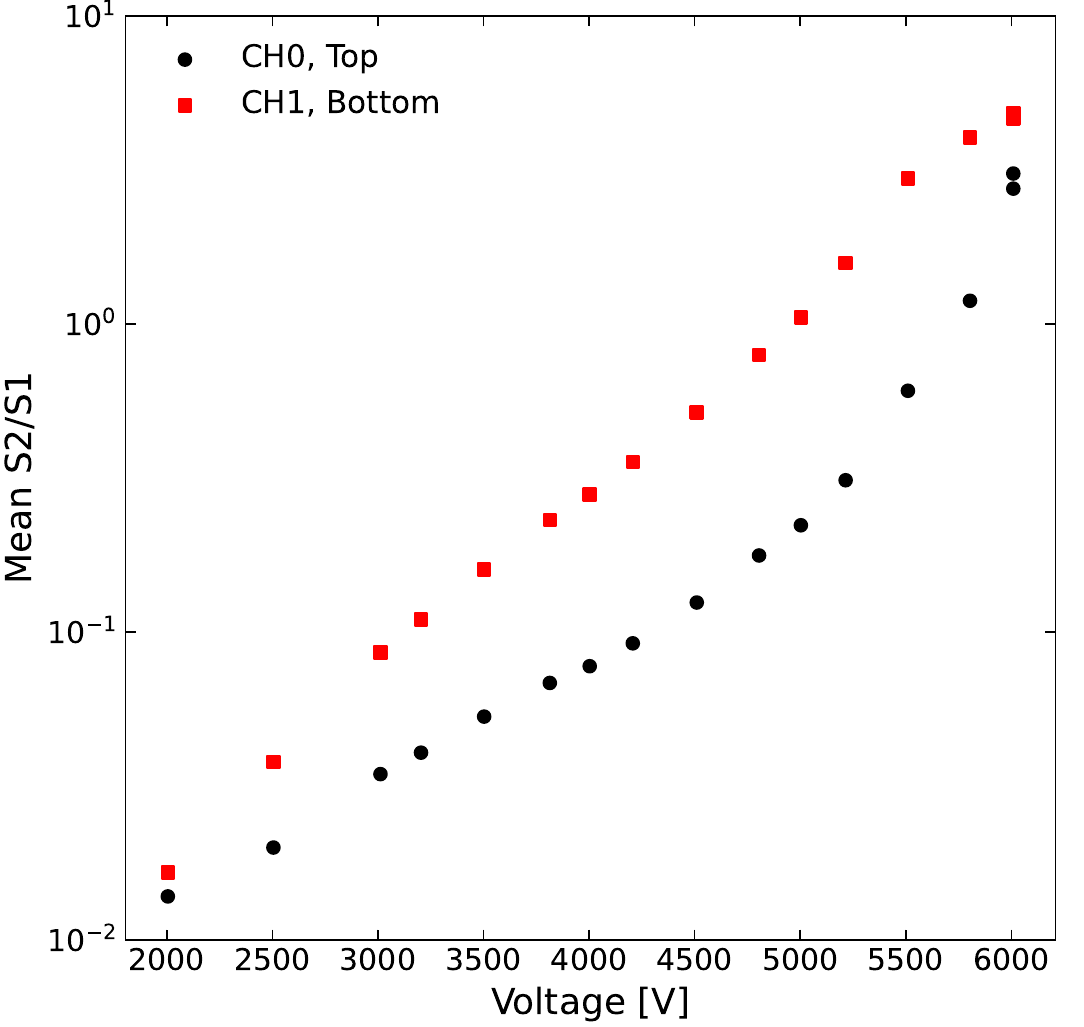}
\caption{Mean of S2/S1 for each run  as a function of needle voltage. The exponential trend expected for proportional amplification is observed. Above $5000\;\si{\volt}$,  signal saturation results in loss of proportionality, which is not observed in the attenuated channels. \label{fig:s2OverS1withVoltage}}
\end{figure}

The S2 signal for data collected at $2505\;\si{\volt}$, $4510\;\si{\volt}$, and $6009\;\si{\volt}$ is shown in Figure~\ref{fig:s2Areas}. 
To measure the change in the amplification gain induced at the needle, the S2 signal for each event was divided by the S1, applying a selection to the data to keep only events with S1 above $10\;\si{\milli\volt \nano\second}$. The distribution of S2/S1 was fit with a Gaussian, and the mean for each run is shown in Figure~\ref{fig:s2OverS1withVoltage} as a function of voltage. 
The value of S2/S1 with voltage follows the expected exponential shape, demonstrating the proportional amplification achieved over a wide range of gains. At $6000\;\si{\volt}$ the FADC was saturated for S2 signals, resulting in deviation in the exponential form.

\begin{figure}[htbp] 
  \centering
\subfigure[\label{subfig:deltat2500}]{\includegraphics[width=0.245\linewidth]{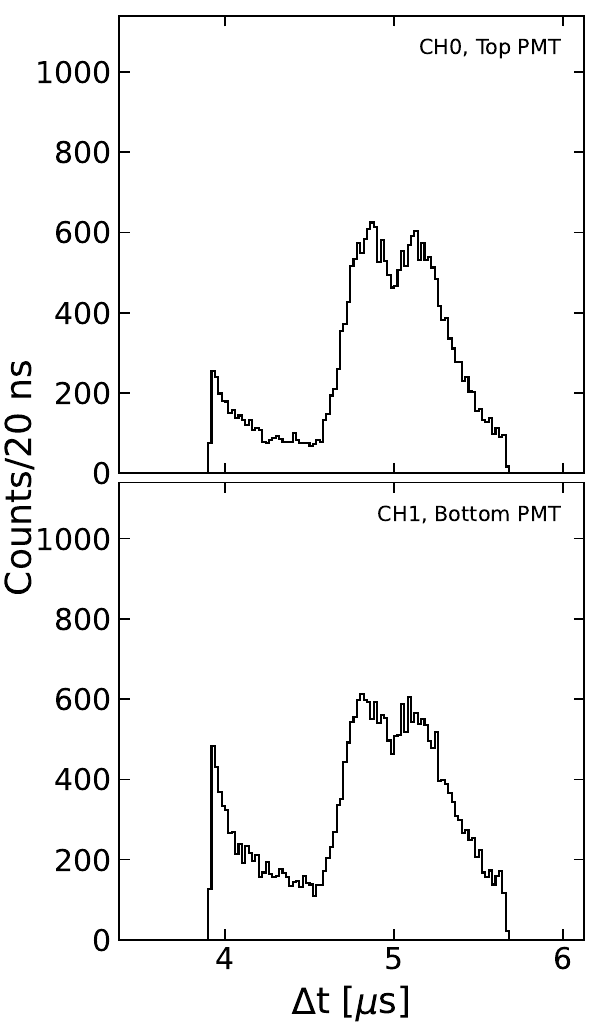}}
\subfigure[\label{subfig:deltat6000}]{\includegraphics[width=0.245\linewidth]{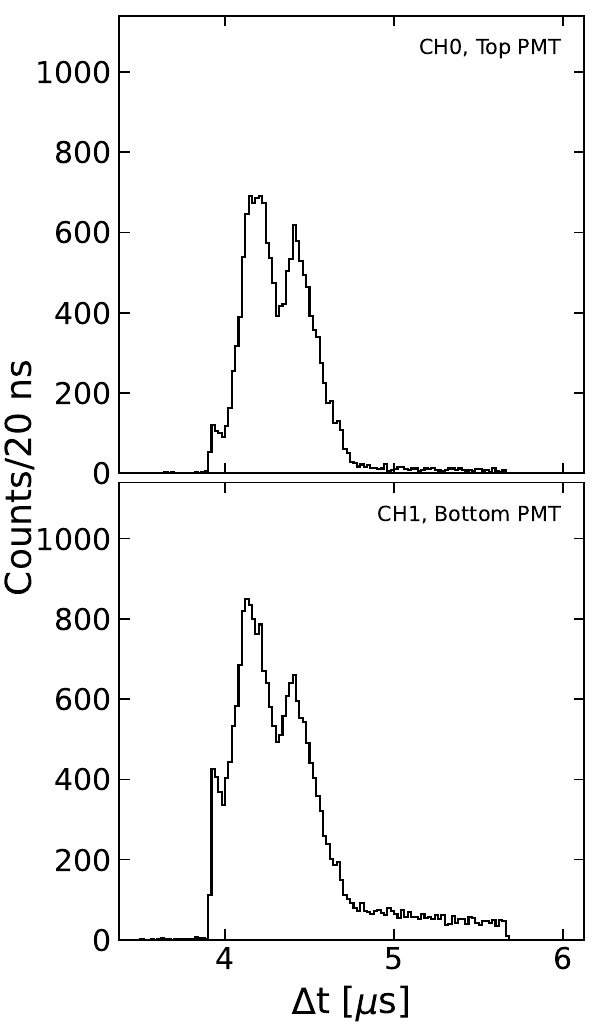}}
\subfigure[\label{subfig:S2oS1vsdeltat2500}]{\includegraphics[width=0.245\linewidth]{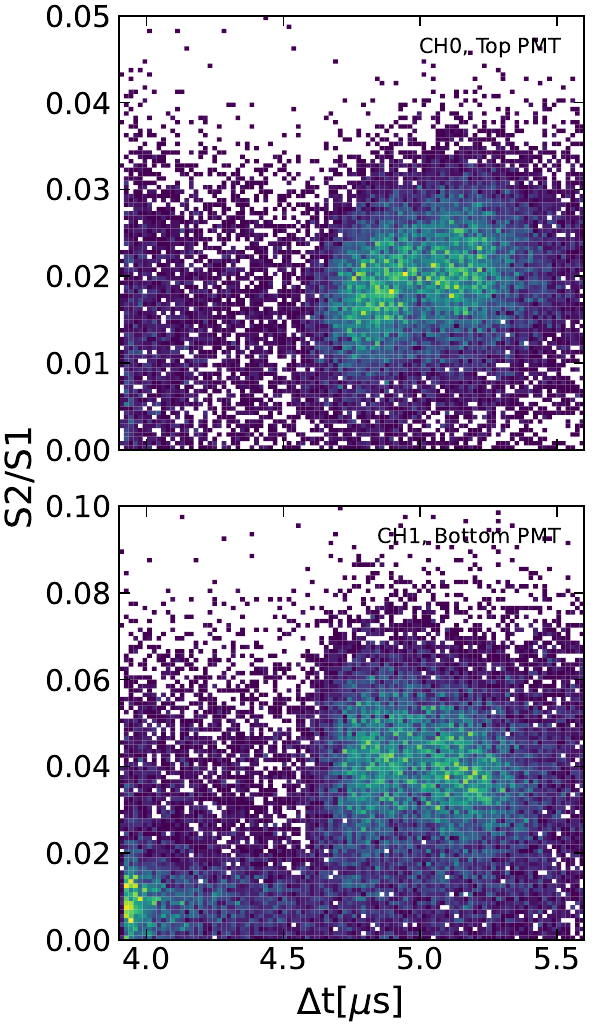}}
\subfigure[\label{subfig:S2oS1vsdeltat6000}]{\includegraphics[width=0.245\linewidth]{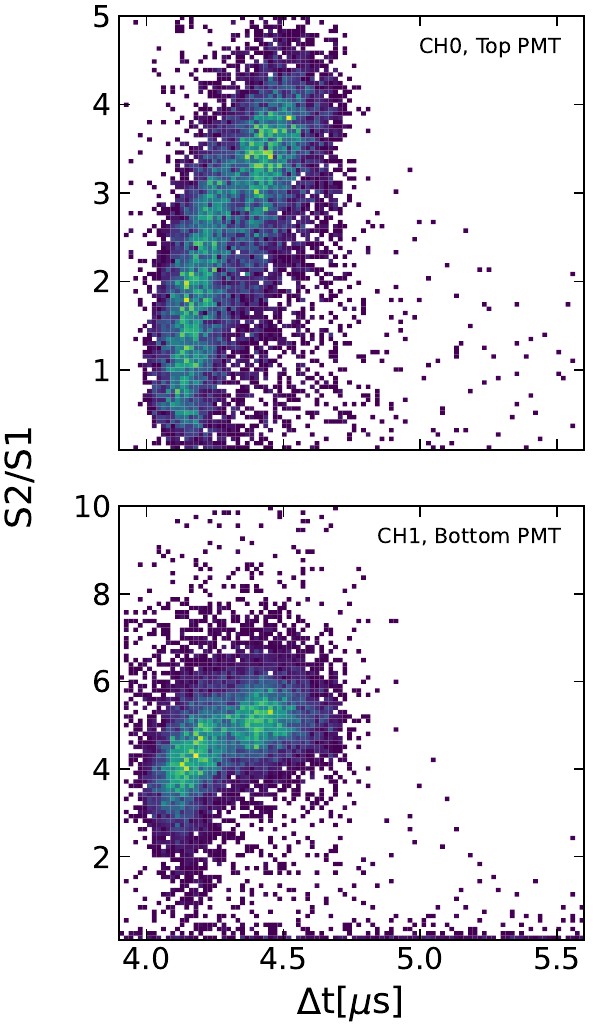}}
  \caption{\label{fig:deltat} $\Delta$t distribution recorded when \subref{subfig:deltat2500} $2505\;\si{\volt}$, \subref{subfig:deltat6000} $6009\;\si{\volt}$ was applied to the needle. S2 to S1 ratio as a function of $\Delta$t for needle voltage of \subref{subfig:S2oS1vsdeltat2500} $2505\;\si{\volt}$ and \subref{subfig:S2oS1vsdeltat6000} $6009\;\si{\volt}$.}
\end{figure}

It was found that the $\Delta t$ decreased with voltage, with examples at $2505\;\si{\volt}$ and $6009\;\si{\volt}$ shown in Figure~\ref{fig:deltat}. This was expected, as the increase in voltage decreases the drift time of ionisation electrons, resulting in a reduced time between the S1 and S2 signals. A double peak structure is observed in the $\Delta t$ distribution for all voltages, which current understanding suggests is of a geometrical origin.

\section{Conclusion} 

Single-phase TPCs present a viable alternative for expanding the capabilities of liquid-phase TPCs in the pursuit of direct DM detection. Utilising charge amplification structures directly in liquid noble elements can significantly enhance sensitivity by improving fiducialization and background discrimination capabilities. However, achieving this requires innovative approaches in designing charge amplification structures, given the high electric fields required. This study presents a novel approach by employing thin, needle-shaped structures to produce a secondary signal (S2) in a single-phase liquid xenon TPC test bench. Initial results indicate that this approach is viable, as evidenced by the successful detection of S2 signals at voltages ranging from $2$ to $6\;\si{\kilo\volt}$. The successful application of needle-shaped structures in producing the secondary signal creates opportunities for progress in DM detection technologies. Further investigations are necessary to improve and optimise the charge amplification structure, explore alternative materials and designs, and expand the application of the technique to larger detectors. This new result can pave the way for enhanced DM detection technologies and is a promising avenue for furthering our comprehension of the composition of DM in the universe.

\acknowledgments

This work was supported by JSPS KAKENHI Grant Numbers JP21K18623, the joint research program of the Institute for Cosmic Ray Research (ICRR), The University of Tokyo, and the  International Exchanges Scheme of the Royal Society, UK (IES\textbackslash R1\textbackslash 211165).

\clearpage
\bibliographystyle{JHEP}
\bibliography{biblio.bib}

\end{document}